\documentclass[final,3p,times,twocolumn]{elsarticle}

\usepackage{amssymb,amsmath,amsthm,bm}
\usepackage{geometry}
\newlength\OneImW
\newlength\ThreeImW
\newlength\vfigskip
\usepackage{graphicx}
\usepackage{url}

\usepackage{subfig}

\usepackage{enumitem}  

\usepackage{mathtools}

\usepackage{IEEEtrantools}

\graphicspath{{figures_pdf/}}

\journal{Journal of Information Security and Applications}
\usepackage{hyperref}

\begin{document}

\begin{frontmatter}

\title{Cryptanalysis of an Image Block Encryption Algorithm Based on Chaotic Maps}

\author{Yunling Ma}

\author{Chengqing Li\corref{corr}}
\ead{DrChengqingLi@gmail.com}

\author{Bo Ou}

\cortext[corr]{Corresponding author.}

\address{College of Computer Science and Electronic Engineering, Hunan University, Changsha 410082, China}

\begin{abstract}
Recently, an image block encryption algorithm was proposed based on some well-known chaotic maps. The authors claim that the encryption algorithm achieves enough security level and high encryption speed at the same time.
In this paper, we give a thorough security analysis on the algorithm from the perspective of modern cryptology and report some critical security defects on the algorithm.
Given five chosen plain-images and the corresponding cipher-images, the attacker can obtain an equivalent secret key to successfully decrypt the other cipher-images encrypted with the same secret key.
In addition, each security metric adopted in the security evaluation on the algorithm is questioned.
The drawn lessons are generally applicable to many other image encryption algorithms.
\end{abstract}
\end{frontmatter}

\section{Introduction}

With the advances in cyberspace technology and rapid popularization of smartphones, people share their photos on various social media platforms with more and more frequency.
As examined in \cite{kim:phyc:TI2017}, such selfie posting behaviors are supported by the complex emotional needs of human.
However, these photos may contain rich privacy information. If they are accessed by some unauthorized person, serious consequences
may happen for the involved parties. Therefore, in the process of image transmission on all kinds of networks, protecting image information from being leaked and stolen has become a focus in the field of information security \cite{ez:private:IS17,ez:verifiable:MIS15}. In recent years, more and more image encryption algorithms have been proposed to withstand the challenge \cite{cqli:meet:JISA19,hua:Diffusion:IEEE19}.

In the past two decades, a large number of image encryption algorithms based on various chaotic systems were proposed
due to their special characteristics, such as sensitivity to the change of initial condition, unpredictability, randomness and high complexity \cite{GZ:cryptographic:IJBC06,cqli:Diode:TCASI19,Wangch:hyper:SP20,Nestor:4D:IS20}.
For example, Hua's encryption algorithm proposed in \cite{zy:Logistic-adjusted-Sine:IS16} uses 2-D Logistic-adjusted-sine map
as a pseudorandom number generator (PRNG); the encryption scheme in \cite{Wang:PRNG:IJBC2019} adopts a 4-D piecewise logistic map with coupled parameters as
the source of pseudorandom number to control all encryption operations; the method designed in \cite{Khan:DNA:IFS19} utilizes a quantum chaotic map as PRNG;
the image encryption system designed in \cite{Ismail:fractional:SP2020} employs four variants of Logistic map as PRNG.
However, in the design of image encryption algorithm based on chaotic systems, many designers ignored another side of chaos:
the order hidden in any chaotic system \cite{Harry:ergodic:MST67}, obvious dynamics degradation of any chaotic system in the domain generated by a digital computer \cite{Yql:degradation:IJBC17,cqli:autoblock:IEEEM18,cql:Dynamic:ITCS19}.
The second point may cause the invalidity of chaotic domain, diverse security pitfalls of many insecure encryption algorithms were reported \cite{GZ:cryptographic:IJBC06,Li:Cryptanalysis:IM2018}.
It can be seen that a chaos-based encryption algorithm may own several special security defects that do not exist in the non-chaotic encryption methods \cite{GZ:cryptographic:IJBC06,F:nonlinear:ND18,cxz:RT-enhanced:IEEEA18}.
As reviewed in \cite{cqli:meet:JISA19}, most chaotic encryption schemes does not target a specific application scenario as \cite{Latif:Quantum:TNSM20}, which provides
a solution for securing sensitive data in the scenario of 5G-based Internet of Things.

In \cite{lfl:chaotic-maps:IET17}, a chaotic image encryption algorithm was proposed,
where three different chaotic maps are used as a PRNG for controlling pixel shuffling, blocking size, and value encryption.
A simple chaotic map is used to determine block size and generate pseudo-random binary sequence for each block.
Especially, the sum of pixels of a plain-image is used to build up a sensitivity mechanism of the encryption result on the plain-image. However,
we found that the security defects of the chaos-based PRNG and canceled the sensitivity mechanism. As for one round version of the algorithm,
we can derive the secret-key with a chosen-plaintext attack. In addition, each used security metric is questioned from the viewpoint of  modern cryptanalysis.

The rest of the paper is organized as follows. Section~\ref{sec:IBEA} presents a description of the analyzed algorithm.
Detailed cryptanalytic results are given in Sec.~\ref{sec:cryptanalysis} with some examples.
The last section concludes the paper.

\section{The image block encryption algorithm based on three chaotic maps}
\label{sec:IBEA}

Assume that the gray-scale plain-image is denoted as a matrix $\bm{I}$ of size $M\times N$. As specified in \cite{lfl:chaotic-maps:IET17},
three chaotic maps are used, namely Arnold map, Baker map, and Logistic map. They are used for permuting position of plain-image, dividing the plain-image into four blocks, and generating pseudo-random number sequences $\{ t_p \}$ and $\{m_p\}$, respectively. The encryption algorithm can be described as follows.

\setlist[itemize]{leftmargin=*}

\begin{itemize}
\item \textit{The secret key}: integers $a, b\in \{0, 1, \cdots, N-1 \}$,
four floating-point numbers $x_0, y_0, z_0, \mu \in [0, 1]$,
floating-point number $r\in (3.5699, 4]$, integer $s_0\in [0, 255]$.

\item \textit{The encryption procedure}:

\textit{Step 1:} Calculating block size with Baker map
\begin{IEEEeqnarray}{rCl}
	\IEEEeqnarraymulticol{3}{l}{(x_{i+1}, y_{i+1})}
	\nonumber\\* \quad
	& = &
	\begin{cases}
		(x_i/\mu,\mu \cdot y_i)                  & \mbox{if } 0<x\leq\mu; \\
		((x_i-\mu)/\mu^*, \mu^*y_i+\mu)  & \mbox{if } \mu<x\leq 1,
	\end{cases}
	\label{equation:Baker_map}
\end{IEEEeqnarray}
where $\mu$ is the control parameter, $\mu^*=1-\mu$. Set the initial value of Baker map~(\ref{equation:Baker_map}) as $((x_0+\mathit{mean}\{\bm I\}/256) \bmod 1$, $ (y_0+\mathit{mean}\{\bm I\}/256) \bmod 1)$, where $\mathit{mean}\{ \cdot \}$ denotes the mean value of the pixels of plain-image $\bm{I}$, $a \bmod m=r$, and $r=a-m \cdot \lfloor a/m \rfloor$.
Then, iterate the map $n=10^4$ times and obtain $u=\lfloor M\cdot x_n \rfloor$ and $v=\lfloor N\cdot y_n \rfloor$.

\textit{Step 2:} Permute the relative position of every pixel of $\bm{I}$ with the permutation relation determined by Arnold map
\begin{equation}
\begin{pmatrix}
p\\
q
\end{pmatrix}=
\begin{pmatrix}
1 & b \\
a & a\cdot b+1
\end{pmatrix}
\begin{pmatrix}
i\\
j
\end{pmatrix}\bmod N,
\label{equation:Arnold_map}
\end{equation}
where $a$, $b$ and $N$ are parameters; $(i, j)$ represents the position of the pixel at the $i$-th row and the $j$-th column of the plain-image $\bm{I}$;
$(p, q)$ represents the permutation position of $(i, j)$.
Swap the pixel at entry $(0, 0)$ with that at $(u, v)$ and obtain the permuted image $\bm{E}$.

\textit{Step 3:} Divide the permuted image $\bm{E}$ into four sub-images $\bm{E}_1=\bm{E}(1:u; 1:v)$; $\bm{E}_2=\bm{E}(1:u; v+1:N)$; $\bm{E}_3=\bm{E}(u+1:M; 1:v)$; $\bm{E}_4=\bm{E}(u+1:M; v+1:N)$.

\textit{Step 4:} Generate pseudo-random sequence $\{s_p\}_{p=1}^{MN}$ by iterating integer Logistic map
\begin{equation}
g^*(x) = 4\cdot x \cdot (256-x)/256
\label{equation:Logistic_map2}
\end{equation}
from initial condition $s_0$. Then, transform sequence $\{s_p\}$ into another one $\{t_p\}$ by
\begin{equation}
h_1(x)=\left\lfloor x \cdot 10^6 \right\rfloor \bmod 256.
\label{equation:trans_Logistic_map2}
\end{equation}

\textit{Step 5:} Generate pseudo-random sequence $\{z_p\}$ by iterating Logistic map
\begin{equation}
g(x) = r\cdot x \cdot (1-x)
\label{equation:Logistic_map1}
\end{equation}
from initial value
\begin{equation}
z_0(1)=(z_0+\mathit{mean}\{\bm I\}/256) \bmod 1.
\label{eq:z01}
\end{equation}
Then, the sequence is further transformed into an integer sequence $\{m_p\}$ via
\begin{equation}
h_2(x)=\left\lfloor x \cdot 10^5 \right\rfloor \bmod 256.
\label{equation:trans_Logistic_map1}
\end{equation}

\textit{Step 6:} Encrypt the $i$-th row and the $j$-th column pixel of the first sub-image $\bm{E}_1$ by
\begin{equation}
a_{ij}^\ast=
\begin{cases}
(a_{ij}+m_p\cdot t_p) \bmod 256      & \mbox{if } m_p=255; \\
(a_{ij}+(m_p+1)\cdot t_p) \bmod 256  & \mbox{otherwise},
\end{cases}
\label{equation:encrypted}
\end{equation}
where $x \bmod 256=x- 256\cdot \lfloor x/256 \rfloor$,
$p=i\cdot N_1+j$.

\textit{Step 7:} As for the other three sub-images $\bm{E}_2$,
$\bm{E}_3$, $\bm{E}_4$, every pixel is encrypt in the same way shown in
\textit{Step 6}. Note that the initial conditions of Eq.~(\ref{equation:Logistic_map1}) corresponding for the three sub-images are set as $z_0(2)$, $z_0(3)$, and $z_0(4)$, respectively. Their values are set as
\begin{equation*}
\begin{cases}
z_0(2)=(z_0(1)+0.5)    \bmod 1; \\
z_0(3)=(z_0(1)+s_0(2)) \bmod 1; \\
z_0(4)=(z_0(2)+s_0(3)) \bmod 1.
\end{cases}
\end{equation*}	
	
\textit{Step 8:} Combine all the encrypted sub-images into the cipher-image $\bm{I^\ast}$ in the corresponding order in the plain-image.

\item The decryption procedure is the inverse version of the encryption algorithm by using the same secret-key. For example, Equation~(\ref{equation:encrypted}) is replaced by its inversion,
\begin{equation*}
		a_{ij}=
		\begin{cases}
			(a_{ij}^\ast-m_p\cdot t_p ) \bmod 256    & \mbox{if } m_p=255; \\
			(a_{ij}^\ast-(m_p+1)\cdot t_p) \bmod 256 & \mbox{otherwise}.
		\end{cases}
\end{equation*}

\end{itemize}

\begin{figure*}[!htb]
\centering
\includegraphics[width=1.2\OneImW]{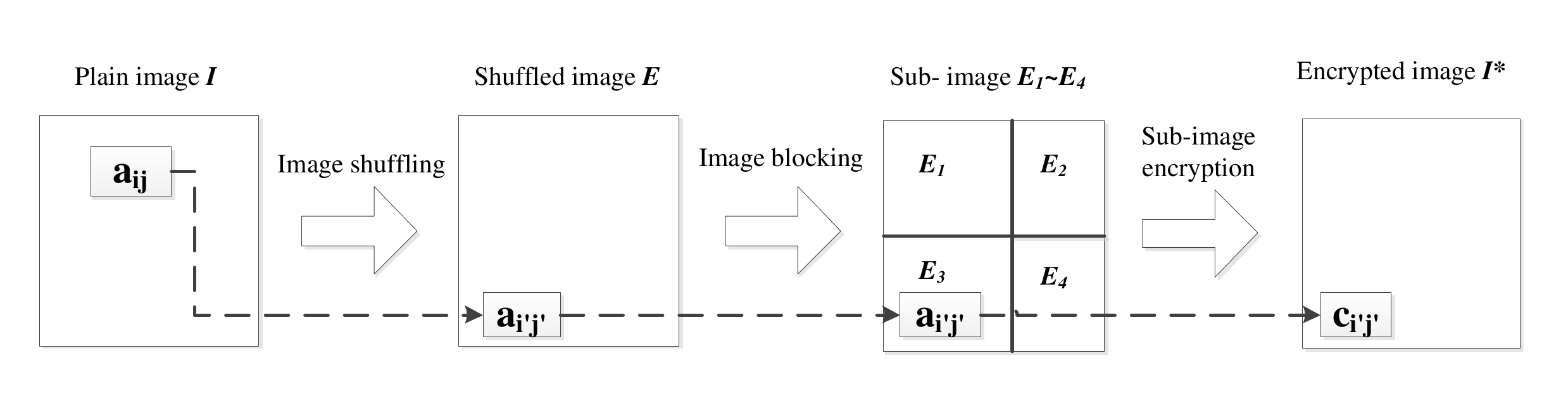}
\caption{The flow chart of the original encryption algorithm.}	
\label{fig:process}
\end{figure*}

\section{Cryptanalysis}	
\label{sec:cryptanalysis}

\subsection{Chosen-plaintext attack}

The original encryption algorithm cascades three parts to encrypt a plain-image: image shuffling, image blocking, and sub-image encryption. The whole flow diagram is shown in Fig.~\ref{fig:process}.

Observing the structure of the analyzed encryption algorithm, one can see that the generation of pseudo-random number sequence is closely related to the mean value of pixels of the plain-image.
If the same secret-key is used, different random sequences will be generated for different mean value of pixels. However, although the plain-images are different, the mean value may be the same, e.g.
some different visible images with the same histogram \cite[Fig. 3]{cql:Inf-Entropy:IEEE18}. In addition, ever for different plain-images with different mean value,
the obtained initial value for the Baker map may be still the same due to the modulo addition with dividend one.
Therefore, as shown in \cite{cqli:autoblock:IEEEM18}, the known-plaintext attack can still work as different plain-images generate the same initial values with a probability larger than 1/256.

Given one pair of plain-image and the corresponding cipher-image, one can try to decrypt another cipher-image if the following two conditions hold at the same time:
(1) the cipher-image is encrypted with the same secret-key ${(x_0, y_0, z_0, r)}$ as the given encrypted image;
(2) the corresponding plain-image of the decrypted cipher-image has the same result calculated with Eq.~(\ref{eq:z01}) as the given plain-image.

To facilitate illustration, we transform Eq.~(\ref{equation:encrypted}) into
\begin{equation}
a_{ij}^\ast=(a_{ij}+P_{ij})\bmod 256,
\label{equation:encrypted2}
\end{equation}
where $\bm{P}=\{P_{ij}\}$ represents the equivalent mask image, namely
\begin{equation}
P_{ij}=
\begin{cases}
(m_p\cdot t_p) \bmod 256      & \mbox{if } m_p=255; \\
((m_p+1)\cdot t_p) \bmod 256  & \mbox{otherwise}.
\end{cases}
\end{equation}

\subsection{Locating the permutation position of the first pixel of the plain-image}
\label{sec:Locating}

Assume that the sum of pixels of the plain-image is denoted as $\eta$. If the generation of random sequence has nothing to do with the sum of pixels of the plain-image,
we know that as long as we use an image with fixed-value zero, we can get an equivalent mask image $\bm{P}$.
However, since the encryption algorithm is related to the sum of pixels of the plain-image, we need to ensure that the sum of pixels is the same as the plain-image,
so we need to adjust the value of the first pixel as $\eta$. There is only one different element between the cipher-image obtained from such an image
and the equivalent mask image $\bm{P}$ of the plain-image $\bm{I}$.
Due to the permutation operation, the different element is located at permutation position of the first pixel in the plain-image $\bm{I}$.
Therefore, we first need to find the permutation position of the first pixel in the plain-image, then obtain the equivalent mask image $\bm{P}$ via Eq.~(\ref{equation:encrypted2}).
So, we choose three plain-images whose sum of pixels are the same as the plain-image.

The permutation position of the first pixel in the plain-image can be revealed by the following steps.

\textit{Step 1: Choose a plain-image $\textbf{q}_1$ of fixed value zero except one pixel $\textbf{q}_1(1, 1)=\eta$} and let $\textbf{Q}_1$ denote the corresponding cipher-image, where
\begin{equation}
	\textbf{q}_1=
	\begin{pmatrix}
		\eta & 0 & 0 & 0 \\
		0 & 0 & 0 & 0 \\
		\vdots & \vdots & \vdots & \vdots \\
		0 & 0 & 0 & 0 \\
		0 & 0 & 0 & 0
	\end{pmatrix}.
\end{equation}

\textit{Step 2: Choose a plain-image $\textbf{q}_2$ of fixed value zero except two pixels $\textbf{q}_2(1, 1)=\eta-1$ and $\textbf{q}_2(1, 2)=1$},
and let $\textbf{Q}_2$ denote the corresponding cipher-image, where
\begin{equation}
	\textbf{q}_2=
	\begin{pmatrix}
		\eta-1 & 1 & 0 & 0 \\
		0 & 0 & 0 & 0 \\
		\vdots & \vdots & \vdots & \vdots \\
		0 & 0 & 0 & 0 \\
		0 & 0 & 0 & 0
	\end{pmatrix}.
\end{equation}

\textit{Step 3: Choose a plain-image $\textbf{q}_3$ of fixed value zero except two pixels $\textbf{q}_3(1, 1)=\eta-1$ and
$\textbf{q}_3(1, 3)=1$}, and let $\textbf{Q}_3$ denote the corresponding cipher-image, where
\begin{equation}
\textbf{q}_3=
	\begin{pmatrix}
		\eta-1 & 0 & 1 & 0 \\
		0 & 0 & 0 & 0 \\
		\vdots & \vdots & \vdots & \vdots \\
		0 & 0 & 0 & 0 \\
		0 & 0 & 0 & 0
	\end{pmatrix}.
\end{equation}

\begin{figure*}[!htb]
	\centering
	\includegraphics[width=\OneImW]{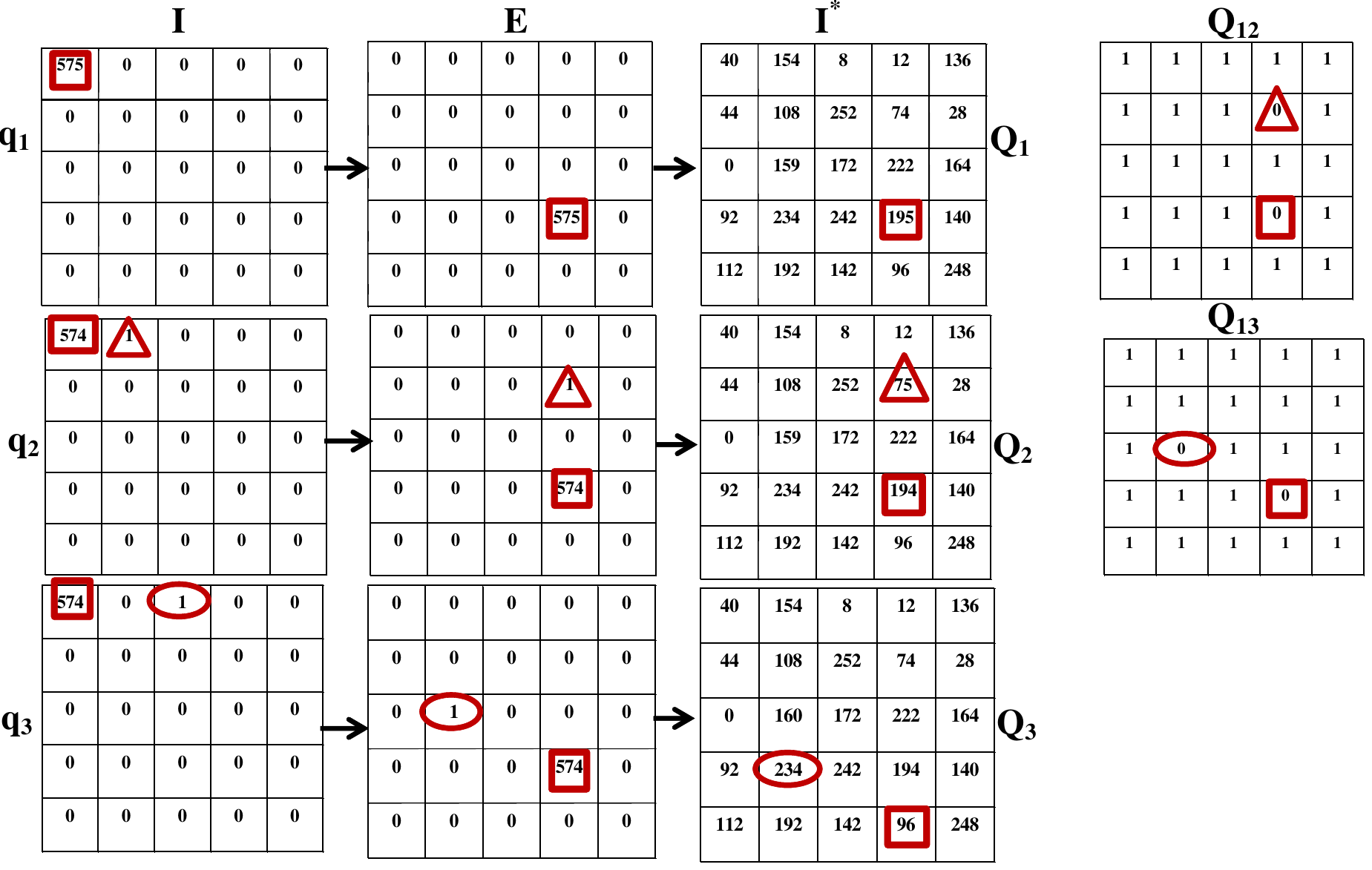}
	\caption{The process of locating the permutation position of the first pixel of the plain-image.}
	\label{fig:first_position}
\end{figure*}

\textit{Step 4: Compare difference between $\textbf{Q}_1$ and $\textbf{Q}_2$.} Let $\textbf{Q}_{12}$ denote the differential image.
One can find that there should be two different pixels between $\textbf{Q}_1$ and $\textbf{Q}_2$. Therefore, there are two elements of value zero in $\textbf{Q}_{12}$, where
\[
\textbf{Q}_{12}(i, j)=0
\]
if $\textbf{Q}_1(i, j)=\textbf{Q}_2(i, j)$.
One can assure that the positions of the two generated zero elements are the permuted positions of two pixels at entries $(1, 1)$ and $(1, 2)$ in the plain-image.

\textit{Step 5: Compare difference between $\textbf{Q}_1$ and $\textbf{Q}_3$}.
Let $\textbf{Q}_{13}$ denote the differential image. One can see that there are two different pixels between $\textbf{Q}_1$ and $\textbf{Q}_3$.
Therefore, there are two elements of value zero in $\textbf{Q}_{13}$, where
$\textbf{Q}_{13}(i, j)=0$
if $\textbf{Q}_1(i, j)=\textbf{Q}_3(i, j)$.
As pointed in the previous analysis, one can see that the positions of the two generated zero elements are the permuted positions of two pixels at entries $(1, 1)$ and $(1, 3)$ in the plain-image.

\textit{Step 6: Compare the locations of Step 4 with Step 5.}
Search for the entry whose value is zero in $\textbf{Q}_{12}$ and $\textbf{Q}_{13}$ at the same time. Then, the obtained position is the permuted position of the pixel at entry $(1, 1)$.

To demonstrate the above process, a concrete example is shown in Fig.~\ref{fig:first_position}, where the sum of pixels of the plain-image is 575, i.e. $\eta=575$, the three chosen plain-images are presented in Fig.~\ref{fig:first_position} in order. As shown in Fig.~\ref{fig:first_position}, the permutation position of the first pixel can be recovered successfully.

\subsection{Recovering the equivalent mask image ${P}$}
\label{sec:shuffle}

From the encryption process, one can see that $\textbf{Q}_1$ is identical to $\bm{P}$ except that $\textbf{Q}_1(u, v)$ as its value is modified. So, subtract $\textbf{Q}_1(u, v)$ with the sum of pixels of the plain-image $(\eta)$, one can derive the equivalent mask image $\bm{P}=\textbf{Q}_1$,
where
\begin{equation}
\textbf{Q}_1(u, v)=(\textbf{Q}_1(u, v)-\eta)\bmod 256.
\end{equation}

Once the equivalent mask image $\bm{P}$ is obtained, the permuted image $\bm{E}$ can be recovered by subtracting the cipher-image $\bm{I}^\ast$ with the equivalent mask image $\bm{P}$,
namely
\begin{equation}
\bm{E}=(\bm{I}^\ast-\bm{P})\bmod 256.	
\end{equation}

\begin{figure*}[!htb]
	\centering
	\includegraphics[width=\OneImW]{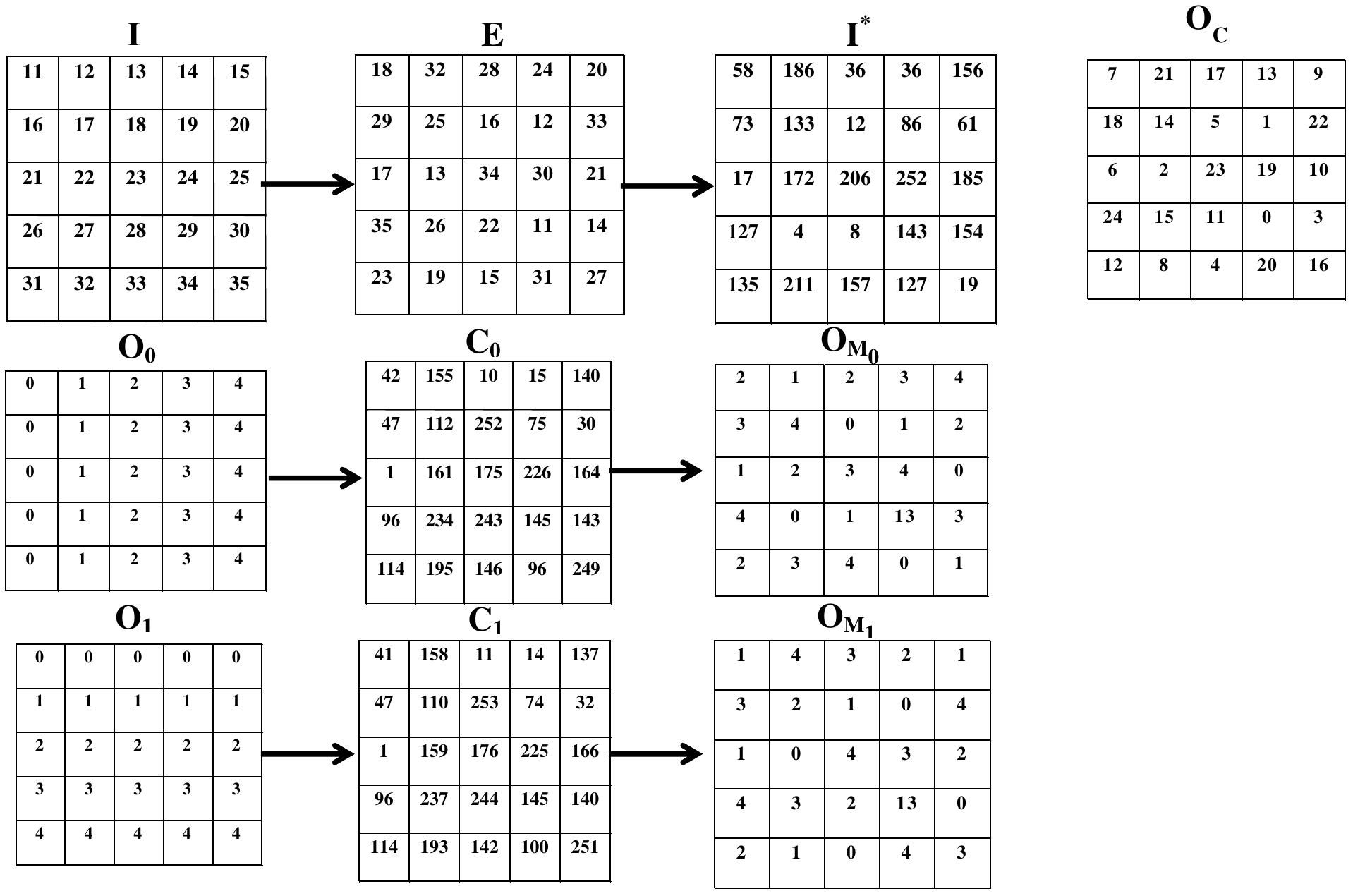}
	\caption{The process of calculating the permutation relation $l$.}
	\label{fig:permutation_rule}
\end{figure*}

\subsection{Recovering the permutation relation generated by Arnold map}

After getting the permuted image, one can use the method in \cite{Lcq:Optimal:SP11,fhj:Cryp-APFM:SP18} to get the secret permutation relation.

If the plain-image $\bm{I}$ is stretched into a vector row by row, the pixel locations become a vector by $\bm{A_0}=\{k\}_{k=0}^{M*N-1}$, which can be expanded to a 2-digit representation with base 256 as follows:
\begin{equation*}
	\textbf{O}=
	\begin{bmatrix}
		(0)(0) & (0)(1) & (0)(2) & \cdots & (0)(255) \\
		(1)(0) & (1)(1) & (1)(2) & \cdots & (1)(255) \\
		\vdots & \vdots & \vdots & \vdots & \vdots  \\
		(254)(0) & (254)(1) & (254)(2) & \cdots & (254)(255) \\
		(255)(0) & (255)(1) & (255)(2) & \cdots & (255)(255)
	\end{bmatrix}_{ M\times N}.
\end{equation*}

Each element in $\textbf{O}$ is a number with base 256. Then, two chosen plain-images with entries $0, 1, 2, \cdots, 254, 255$ are obtained by splitting matrix $\textbf{O}$ into two images
\begin{equation*}
\textbf{O}_0=
	\begin{bmatrix}
		0 & 1 & 2 & \cdots & 255 \\
		0 & 1 & 2 & \cdots & 255 \\
		\vdots & \vdots & \vdots & \vdots & \vdots \\
		0 & 1 & 2 & \cdots & 255 \\
		0 & 1 & 2 & \cdots & 255
	\end{bmatrix}_{M\times N}
\end{equation*}
and
\begin{equation*}
	\textbf{O}_1=
	\begin{bmatrix}
		0 & 0 & 0 & \cdots & 0 \\
		1 & 1 & 1 & \cdots & 1 \\
		\vdots & \vdots & \vdots & \vdots & \vdots \\
		254 & 254 & 254 & \cdots & 254 \\
		255 & 255 & 255 & \cdots & 255
	\end{bmatrix}_{M\times N}.
\end{equation*}

To keep the sum of pixels of these images consistent with the plain-image, as Sec.~\ref{sec:Locating}, the first pixels in two chosen plain-images can be calculated by
\begin{equation*}
	\begin{cases}
	\textbf{O}_{\rm se}=\sum\limits_{i=1}^M\sum\limits_{j=1}^N\textbf{O}_{\rm e}(i,j)-\textbf{O}_{\rm e}(1,1),\\
		\textbf{O}_{\rm e}(1,1)=\eta-\textbf{O}_{\rm se},
	\end{cases}	
\end{equation*}
where $e=(0, 1)$.

The cryptanalysis steps on permutation can be presented as follows:
\begin{itemize}
  \item \textit{Step 1:}  Encrypt two images $\textbf{O}_0$ and $\textbf{O}_1$, and obtain the corresponding cipher-images $\textbf{C}_0$ and $\textbf{C}_1$.

  \item \textit{Step 2:}  Obtain the permuted images $\textbf{O}_{\rm M0}$ and $\textbf{O}_{\rm M1}$ as shown in Sec.~\ref{sec:shuffle}.

  \item \textit{Step 3:}  Derive the permutation rule $l$.
\end{itemize}

First, one can get $\textbf{O}_{\rm c}=\{\textbf{O}_{\rm c}(i,j)\}_{i=1, j=1}^{M, N}$.
Representing in the form of matrix-wise operation, one has
\begin{equation*}
\textbf{O}_{\rm c}=256 \cdot \textbf{O}_{\rm M1}+\textbf{O}_{\rm M0}.
\end{equation*}
Because the first pixels in $\textbf{O}_0$ and $\textbf{O}_1$ are adjusted, the element $\textbf{O}_{\rm c}(u, v)$ is wrong. But, the rest locations are correct; the correct $\textbf{O}_{\rm c}(u, v)$ can be re-calculated by
\begin{equation*}
	\begin{cases}
		\textbf{O}_{\rm sc}=\bigg(\sum\limits_{i=1}^M \sum\limits_{j=1}^N\textbf{O}_{\rm c}(i,j)\bigg)-\textbf{O}_{\rm c}(u, v), \\
		\textbf{O}_{\rm c}(u, v)=\bigg(\sum\limits_{k=0}^{MN-1}k\bigg) -\textbf{O}_{\rm sc}.
	\end{cases}	
\end{equation*}
Therefore, the permutation rule $l$ described by vector $\bm{L}_0$ can be obtained by stretching $\textbf{O}_{\rm c}$ to a vector row by row. Using $\bm{L}_0$, one can obtain the original plain-image $\bm{I}$ from the permuted image $\bm{E}$.

To illustrate the above process, a concrete example is shown in Fig.~\ref{fig:permutation_rule}.
From the figure, one can see that the permutation position of the first pixel is $(4, 4)$, which is consistent with the result calculated in Sec.~\ref{sec:Locating}, which verifies eligibility of the above method. Note that the sum of pixel of the legend image is 575 ($\eta=575$).

\subsection{The deciphering process for a given cipher-image}

Summarizing the above analyses, the decryption process of a given cipher-image using an equivalent secret key can be described as follows.
\begin{itemize}
  \item \textit{Step 1}: Locate the permutation position of the first pixel of the plain-image.
  \item \textit{Step 2}: Recover the equivalent mask image $\bm{P}$.
  \item \textit{Step 3}: Get the permuted image.
  \item \textit{Step 4}: Obtain the plain-image.
\end{itemize}

\begin{figure*}[!htb]
\centering
\begin{minipage}{\ThreeImW}
\centering
\includegraphics[width=\ThreeImW]{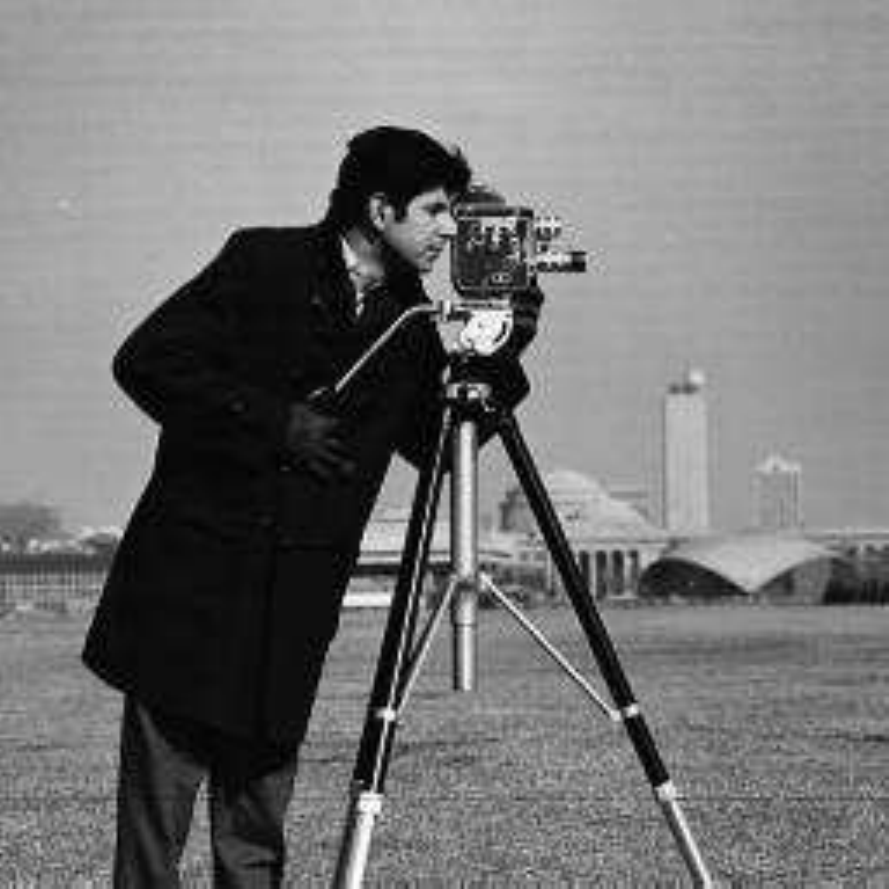}\\
(a)
\end{minipage}\hspace{\vfigskip}
\begin{minipage}{\ThreeImW}
\centering
\includegraphics[width=\ThreeImW]{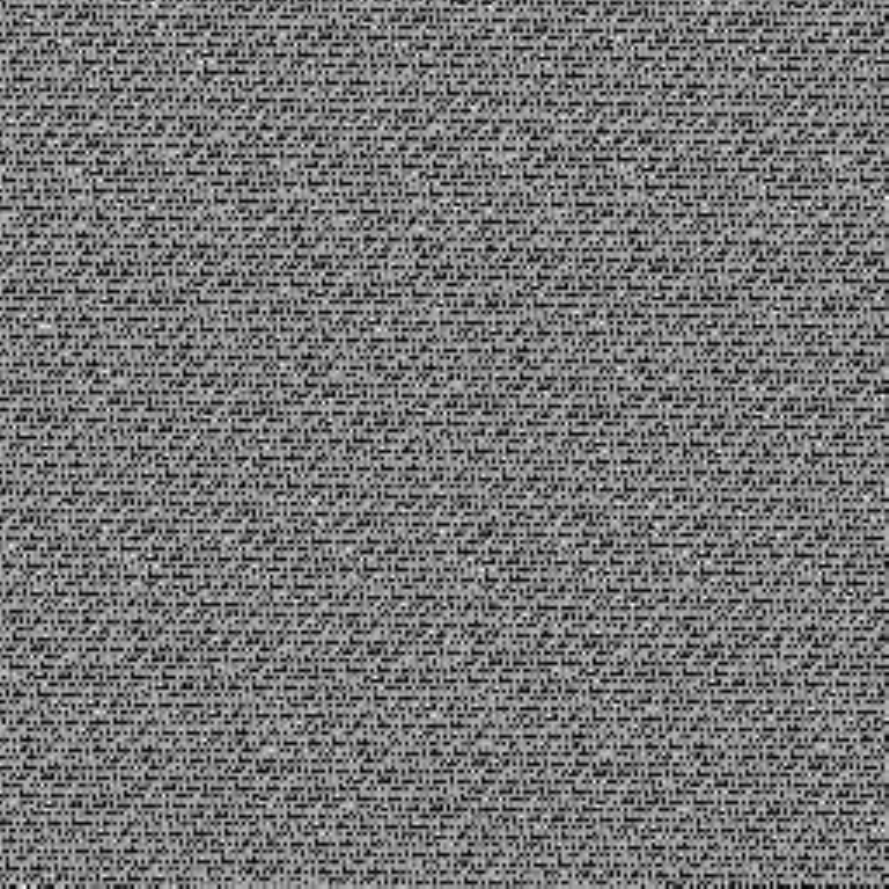}\\
(b)
\end{minipage}\hspace{\vfigskip}
\begin{minipage}{\ThreeImW}
\centering
\includegraphics[width=\textwidth]{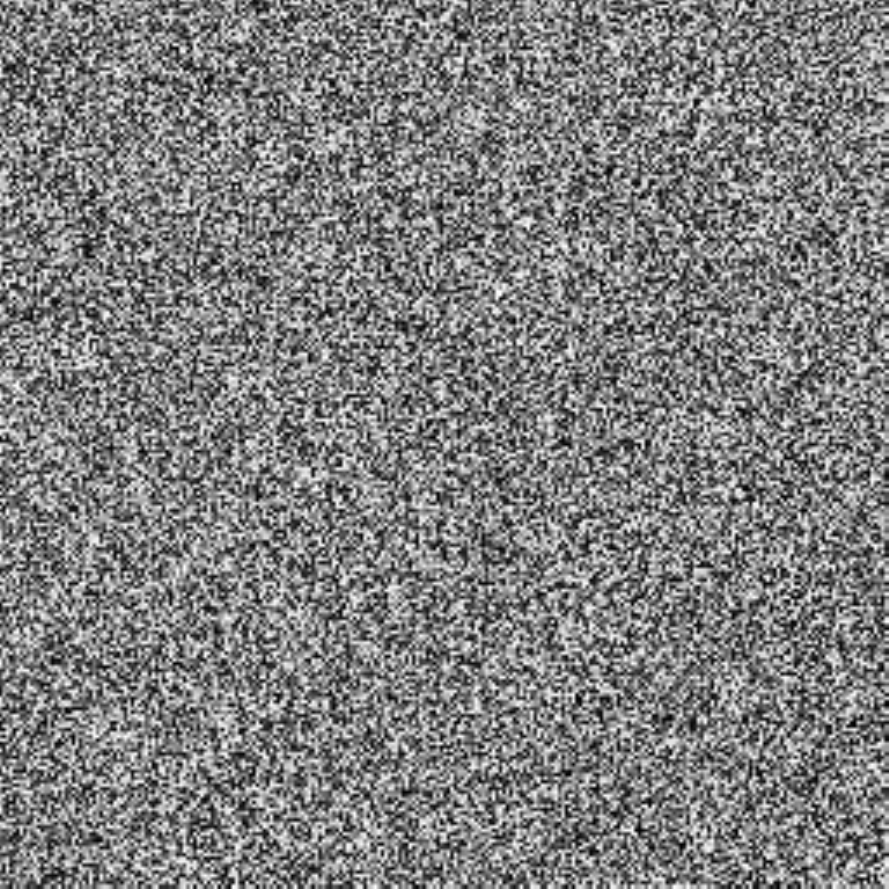}
(c)
\end{minipage}
\caption{The images in the encryption process:
(a) The original-image;
(b) The permuted-image;
(c) The cipher-image}
\label{figure:encryption}	
\end{figure*}

\begin{figure*}[!htb]
\centering
\begin{minipage}{\ThreeImW}
\centering
\includegraphics[width=\ThreeImW]{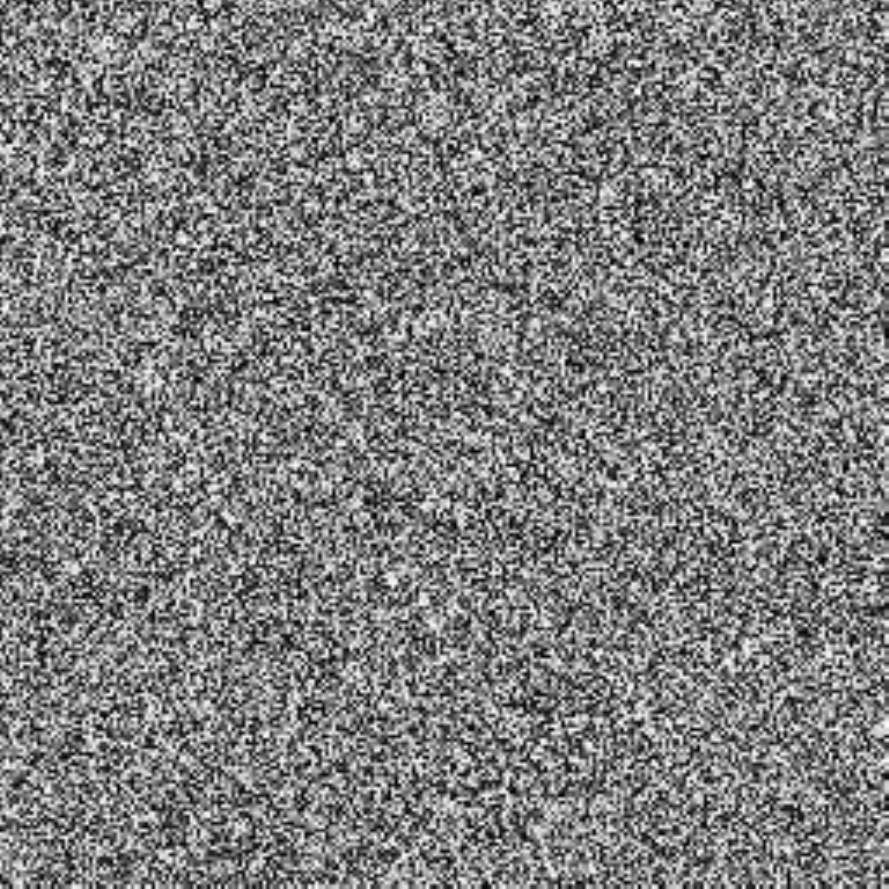}
(a)
\end{minipage}\hspace{\vfigskip}
\begin{minipage}{\ThreeImW}
\centering
\includegraphics[width=\ThreeImW]{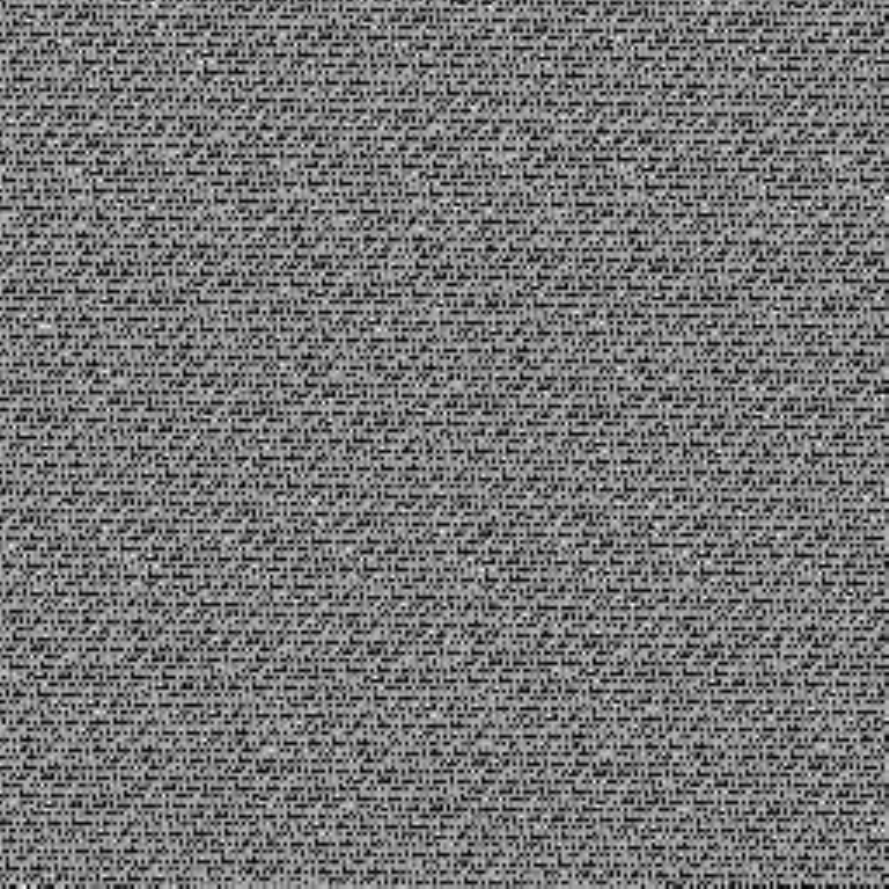}
(b)
\end{minipage}\hspace{\vfigskip}
\begin{minipage}{\ThreeImW}
\centering
\includegraphics[width=\ThreeImW]{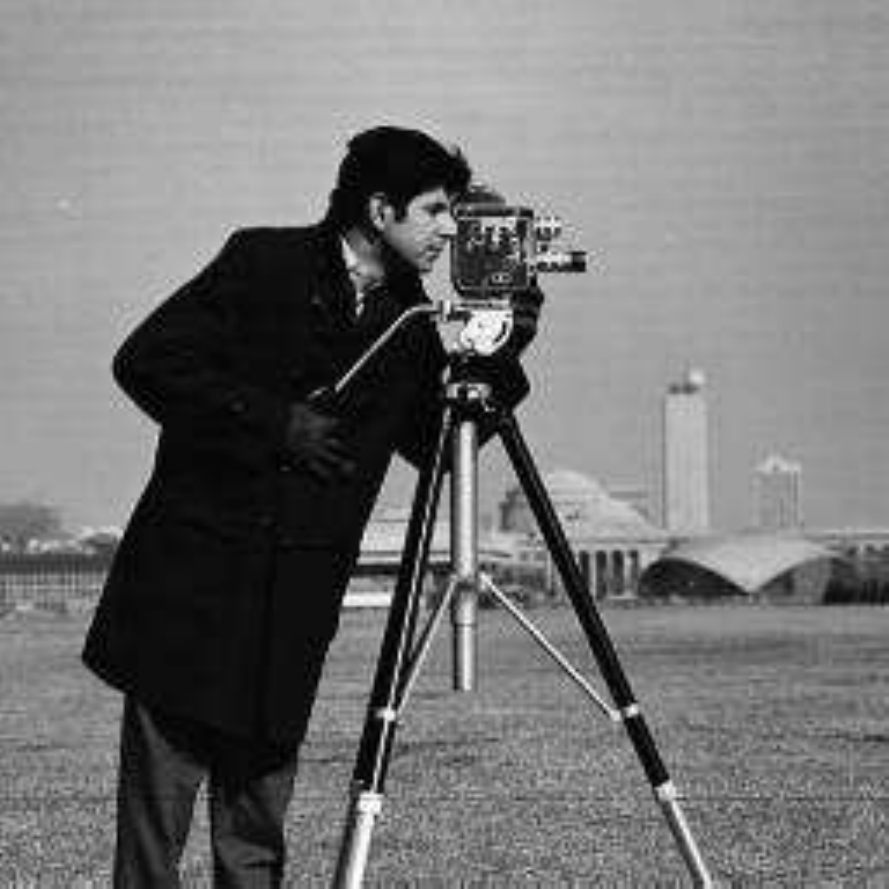}
(c)
\end{minipage}
\caption{The results of chosen-plaintext attack:
(a) recovered mask image $\bm{P}$;
(b) recovered permuted image;
(c) decrypted plain-image.}
\label{figure:deciphering}	
\end{figure*}

To further check the overall performance of  the proposed attacking method, we performed a number of experiments with some typical secret keys. When the secret keys and parameter are set as \cite{lfl:chaotic-maps:IET17}, namely $x_0=0.123$, $y_0=0.456, n=10000$, $s_0(1)=0.789, z_0=0.147$, $a=97$, $b=111$ and $r=3.999$.
The plain-image ``Cameraman" and the encryption results are shown in Fig.~\ref{figure:encryption}.
and Fig.~\ref{figure:deciphering}.

\setlength\abovecaptionskip{1pt}

\subsection{Security defects of the encryption algorithm under study}

Like most recent papers on designing image encryption algorithms, reference \cite{lfl:chaotic-maps:IET17} uses the general security evaluation criteria used in them to
prove the security of the proposed encryption algorithm.
However, we try to prove that the adopted criteria are not proper in terms of security evaluation.
Here, we try to rebut one by one according to the order of security analysis in the original paper \cite{lfl:chaotic-maps:IET17}.
\begin{itemize}
\item{Histogram analysis}

In \cite{lfl:chaotic-maps:IET17}, it is stated that ``The readability of image would be less effective with the smoother histogram distribution". As shown in \cite{cqli:autoblock:IEEEM18},
although attackers can not get some meaningful information from the uniform histogram in terms of pixel, they can restore some important statistical information about the plain-image
by changing the counting objects of the histogram from pixel to bit. In addition, the secret permutation keeping histogram unchanged is also
efficient for some application scenarios needs lightweight cryptography, e.g. pay-TV, RFID, which is emphasized in \cite{cql:hierarchical:SP16}.
Therefore, the two histograms calculated in terms of pixel shown in \cite{lfl:chaotic-maps:IET17} are far not enough to prove that the proposed encryption algorithm has
satisfying security performance.

\item{Variance of histogram}

This criteria evaluates the uniformity of a cipher-image by calculating the variance of histogram. In \cite{lfl:chaotic-maps:IET17}, it is said that
``The lower value of variances indicates the higher uniformity of encrypted images. Comparing with the variance of histogram of the plaintext image,
we have that the proposed algorithm is efficient and can resist such statistical attacks effectively".
In fact, the variance of histogram can not measure the number of different pixel value from the histogram of a cipher-image.
For example, the variances of two histograms ``2, 2, 3, 3, 4, 7" and ``2, 2, 3, 3, 5, 6" are different, but their number of different combinations are the same.
Besides, reference \cite{Preishuber:motivation:TIFS2018} prove that a cipher-image generated by an insecure encryption algorithm can also obtain very low
variance of histogram.

\item{Key space analysis}

In \cite{lfl:chaotic-maps:IET17}, the precision of the secret key is fixed as $10^{-14}$. However, as shown in \cite{cql:Inf-Entropy:IEEE18}, in digital computer, the precision can only be precisely specified by a power of two. If data is represented in floating-point format binary32 or binary64, the distance between adjacent representable numbers is not even, so the length of mantissa fraction and exponential fraction need to be set carefully \cite{cql:Dynamic:ITCS19}.
Therefore, it is groundless to claim that there are very rich secret keys to resist different kinds of brute-force attacks.
As analysis on the state-mapping network of Cat map implemented in digital computer in \cite{cqli:cat:2019},
there are a number of states resulting in very short period \cite{Fan:period:ND19}, which means that a number of weak secret key should be removed.

\item{Key sensitivity analysis}

In \cite{lfl:chaotic-maps:IET17}, the impact of the so-called small changes of the secret key on the encryption and decryption process is investigated by changing one of its five keys with fluctuation up to $10^{-14}$. However, in the computer, any calculation is based on binary system, so the sensitivity of secret keys is not credible by adopting a change presented by a decimal system.
For example, $0.1=(0.0 0011 0011 0011 0011 0011\cdots)_2$. So a seemingly small number in terms of decimal base may mean substantial change in terms of binary presentation.
In addition, observing Eq.~(\ref{equation:Logistic_map2}) and (\ref{equation:Logistic_map1}), one can see that if Logistic map is implemented in a fixed-point arithmetic domain,
$x$ is equivalent to $(256-x)$ and $(1-x)$, respectively.

Due to the modulo addition in Eq.~(\ref{equation:trans_Logistic_map1}), there may exist even much more equivalent secret keys.
Besides these, the quantization effect of digital chaotic maps may incur the same iteration orbit for different initial conditions.
Therefore, the sensitivity of the encryption results with respect to change of secret key is weak.

\item{Correlation analysis}

In \cite{lfl:chaotic-maps:IET17}, a popular measurement method is used to calculate the correlation degree of adjacent pixels in the horizontal direction, vertical direction and diagonal direction.
It is claimed that ``There is no clear (statistic) decision criterion for passing this test."
The authors of  \cite{lfl:chaotic-maps:IET17} believe that the ideal value should be close to 0. So the lower the calculated value between adjacent pixels, the better security performance of the encryption algorithm.
In \cite{Preishuber:motivation:TIFS2018}, several insecure encryption algorithms are deliberately constructed to calculate the correlation among adjacent pixels in the
three directions, and the value is also close to 0, which means that the test results are not sufficient but only necessary  for the expected performance.

\item Information entropy analysis

Information entropy is used to measure the ``randomness" of a cipher-image and the idea best value is 8.
As demonstrated by some counter-examples given in \cite{Preishuber:motivation:TIFS2018}, some encryption algorithms with very poor security performance
can also produce cipher-images with information entropy of value 8.
Therefore, passing the test is only a necessary condition, but not a sufficient one for securing a high security level.

\item Resistance to differential attack analysis

Two parameters $\mathit{NPCR}$ and $\mathit{UACI}$ are used to evaluate the sensitivity to the change of plain-image.
Using the result approaches the ideal value, they claimed that the encryption result of the encryption algorithm under study is extremely sensitive to change of plain-image and can resist differential attack effectively.
However, the values for four insecure ciphers in \cite{Preishuber:motivation:TIFS2018} are above 99\%, which is comparable to the best results on the two parameters in the literatures on chaos-based ciphers.
In particular, the values on the two parameters virtually identical to that of some chaos-based encryption schemes reported insecure in \cite{F:nonlinear:ND18}.
As a conclusion, this criterion for measuring encryption security is not credible.

\item{Robustness analysis}

In \cite{lfl:chaotic-maps:IET17}, different levels of noise and data loss are used to verify the robustness of the algorithm under study.
Actually, the statement ``decrypted image can still be recognised clearly under the influence of noise and data loss" relies on the following facts:
(1) human eyes can tolerate serious image noises and extracting significant features;
(2) there is strong correlation in natural images, i.e. the neighbouring pixels are close in terms of intensity value with a non-negligible probability;
(3) as for a symmetric-key encryption scheme, weak sensitivity of decrypted result with respect to change of cipher-image actually means
weak sensitivity of encryption result concerning the change of plain-image.
In addition, we find that there is a serious bug. Before decryption, the receiver needs to obtain the transmitted mean value of pixels of the plain-image accurately.
As shown in the part \textit{key space analysis}, the plain-image can not be decrypted correctly if only one bit of the mean value is wrong in many cases.
Therefore, the bits representing the mean value should be secretly sent to the receivers.

\item{Algorithm speed analysis}

In \cite{lfl:chaotic-maps:IET17}, by comparing several different encryption algorithms, the author gets the conclusion that their algorithm has the best operation speed. Actually, the fast speed is build on sacrificing security instead of better structure. In Eq.~(\ref{equation:trans_Logistic_map1}) and (\ref{equation:trans_Logistic_map2}), authors use the general integer conversion functions, namely
\begin{equation}
f_n(x)=f(10^m \cdot x) \bmod D,
\end{equation}
where $m$ and $D$ are positive integers; $f(x)$ is an quantization function, i.e. ceil, round and floor \cite{cql:Inf-Entropy:IEEE18}. In digital computer, constant multiplication is performed by a series of bitwise shifts and addition operations. Therefore, the computational complexity of the conversion is proportional to $m$. Only $\lceil \log_2{D} \rceil$ bits are useful for encryption, the other $m{\left\lceil \log_2{10} \right\rceil}-{\left\lceil \log_2{D} \right\rceil}$ bits are wasted. Taking Eq.~(\ref{equation:trans_Logistic_map1}) as an example, the utilization percentage of the computation cost on iterating Logistic map (\ref{equation:Logistic_map1}) is only $\frac{\left\lceil \log_2{D} \right\rceil}{m{\left\lceil \log_2{10} \right\rceil}}=\frac{\left\lceil \log_2{256} \right\rceil}{5{\left\lceil \log_2{10} \right\rceil}}=\frac{2}{5}$.
\end{itemize}

\section{Conclusion}

This paper analyzed the security of an image block encryption algorithm based on multiple chaotic maps. Its equivalent secret key can be easily recovered with some chosen plain-images.
Every used security metric is incapable to check its real security performance. The summarized security defects can be used to inform designers of image encryption algorithms about common
security pitfalls in the field of image security, especially chaotic cryptography. Besides, much cryptanalytic works should be done to bridge the gap among nonlinear sciences, signal processing, and cryptography.

\section*{Acknowledgement}

This research was supported by the Natural Science Foundation of China (No.~61772447), the Joint Funds of the National Natural Science Foundation of China, China General Technology Research Institute (No.~U1736113).


\bibliographystyle{elsarticle-num}
\bibliography{IBEA}

\begin{thebibliography}{10}
\expandafter\ifx\csname url\endcsname\relax
  \def\url#1{\texttt{#1}}\fi
\expandafter\ifx\csname urlprefix\endcsname\relax\def\urlprefix{URL }\fi
\expandafter\ifx\csname href\endcsname\relax
  \def\href#1#2{#2} \def\path#1{#1}\fi

\bibitem{kim:phyc:TI2017}
J.~W. Kim, M.~Chock, Personality traits and psychological motivations
  predicting selfie posting behaviors on social networking sites, Telematics
  and Informatics 34~(5) (2017) 560--571.
\newblock \href {http://dx.doi.org/10.1016/j.tele.2016.11.006}
  {\path{doi:10.1016/j.tele.2016.11.006}}.

\bibitem{ez:private:IS17}
E.~Zhang, F.~Li, B.~Niu, Y.~Wang, Server-aided private set intersection based
  on reputation, Information Sciences 387 (2017) 180--194.
\newblock \href {http://dx.doi.org/10.1016/j.ins.2016.09.056}
  {\path{doi:10.1016/j.ins.2016.09.056}}.

\bibitem{ez:verifiable:MIS15}
E.~Zhang, P.~Yuan, J.~Du, Verifiable rational secret sharing scheme in mobile
  networks, Mobile Information Systems 2015 (2015) art. no. 462345.
\newblock \href {http://dx.doi.org/10.1155/2015/462345}
  {\path{doi:10.1155/2015/462345}}.

\bibitem{cqli:meet:JISA19}
C.~Li, Y.~Zhang, E.~Y. Xie, {When an attacker meets a cipher-image in 2018: A
  Year in Review}, Journal of Information Security and Applications 48 (2019)
  art. no. 102361.
\newblock \href {http://dx.doi.org/10.1016/j.jisa.2019.102361}
  {\path{doi:10.1016/j.jisa.2019.102361}}.

\bibitem{hua:Diffusion:IEEE19}
Z.~Hua, B.~Xu, F.~Jin, H.~Huang, Image encryption using josephus problem and
  filtering diffusion, IEEE Access 7 (2019) 8660--8674.
\newblock \href {http://dx.doi.org/10.1109/ACCESS.2018.2890116}
  {\path{doi:10.1109/ACCESS.2018.2890116}}.

\bibitem{GZ:cryptographic:IJBC06}
G.~\'{A}lvarez, S.~Li, Some basic cryptographic requirements for chaos-based
  cryptosystems, International Journal of Bifurcation and Chaos 16~(8) (2006)
  2129--2151.
\newblock \href {http://dx.doi.org/10.1142/S0218127406015970}
  {\path{doi:10.1142/S0218127406015970}}.

\bibitem{cqli:Diode:TCASI19}
N.~Wang, C.~Li, H.~Bao, M.~Chen, B.~Bao, Generating multi-scroll {C}hua's
  attractors via simplified piecewise-linear {C}hua's diode, IEEE Transactions
  on Circuits and Systems I: Regular Papers 66~(12) (2019) 4767--4779.
\newblock \href {http://dx.doi.org/10.1109/TCSI.2019.2933365}
  {\path{doi:10.1109/TCSI.2019.2933365}}.

\bibitem{Wangch:hyper:SP20}
M.~Zhou, C.~Wang, A novel image encryption scheme based on conservative
  hyperchaotic system and closed-loop diffusion between blocks, Signal
  Processing 171 (2020) art. no. 107484.
\newblock \href {http://dx.doi.org/10.1016/j.sigpro.2020.107484}
  {\path{doi:10.1016/j.sigpro.2020.107484}}.

\bibitem{Nestor:4D:IS20}
N.~Tsafack, J.~Kengne, B.~A.-E.-A.~A. M.Iliyasu, KaoruHirota, A.~A.
  EL-Latifchi, Design and implementation of a simple dynamical {4-D} chaotic
  circuit with applications in image encryption, Information Sciences 515
  (2020) 191--217.
\newblock \href {http://dx.doi.org/10.1016/j.ins.2019.10.070}
  {\path{doi:10.1016/j.ins.2019.10.070}}.

\bibitem{zy:Logistic-adjusted-Sine:IS16}
Z.~Hua, Y.~Zhou, {Image encryption using 2D Logistic-adjusted-Sine map},
  Information Sciences 339 (2016) 237--253.
\newblock \href {http://dx.doi.org/10.1016/j.ins.2016.01.017}
  {\path{doi:10.1016/j.ins.2016.01.017}}.

\bibitem{Wang:PRNG:IJBC2019}
Y.~Wang, Z.~Zhang, G.~Wang, D.~Liu, A pseudorandom number generator based on a
  4{D} piecewise logistic map with coupled parameters, International Journal of
  Bifurcation and Chaos 29~(9) (2019) art. no. 1950124.
\newblock \href {http://dx.doi.org/10.1142/S0218127419501244}
  {\path{doi:10.1142/S0218127419501244}}.

\bibitem{Khan:DNA:IFS19}
J.~S. Khan, J.~Ahmad, S.~S. Ahmed, H.~A. Siddiqa, S.~F. Abbasi, S.~K. Kayhan,
  {DNA} key based visual chaotic image encryption, Journal of Intelligent
  \textup{\&} Fuzzy Systems 37~(2) (2019) 2549--2561.
\newblock \href {http://dx.doi.org/10.3233/JIFS-182778}
  {\path{doi:10.3233/JIFS-182778}}.

\bibitem{Ismail:fractional:SP2020}
S.~M. Ismail, L.~A. Said, A.~G.Radwan, A.~H.Madian, M.~F. Abu-ElYazeed, A novel
  image encryption system merging fractional-order edge detection and
  generalized chaotic maps, Signal Processing 167 (2020) 107280.
\newblock \href {http://dx.doi.org/10.1016/j.sigpro.2019.107280}
  {\path{doi:10.1016/j.sigpro.2019.107280}}.

\bibitem{Harry:ergodic:MST67}
H.~Furstenberg, Disjointness in ergodic theory, minimal sets, and a problem in
  diophantine approximation, Mathematical Systems Theory 1 (1967) 1--49.
\newblock \href {http://dx.doi.org/10.1007/BF01692494}
  {\path{doi:10.1007/BF01692494}}.

\bibitem{Yql:degradation:IJBC17}
Y.~Liu, Y.~Luo, S.~Song, Counteracting dynamical degradation of digital chaotic
  chebyshev map via perturbation, International Journal of Bifurcation and
  Chaos 27 (2017) art. no. 1750033.
\newblock \href {http://dx.doi.org/10.1142/S021812741750033X}
  {\path{doi:10.1142/S021812741750033X}}.

\bibitem{cqli:autoblock:IEEEM18}
C.~Li, D.~Lin, J.~L\"u, F.~Hao, Cryptanalyzing an image encryption algorithm
  based on autoblocking and electrocardiography, IEEE MultiMedia 25~(4) (2018)
  46--56.
\newblock \href {http://dx.doi.org/10.1109/MMUL.2018.2873472}
  {\path{doi:10.1109/MMUL.2018.2873472}}.

\bibitem{cql:Dynamic:ITCS19}
C.~Li, B.~Feng, S.~Li, J.~Kurths, G.~Chen, Dynamic analysis of digital chaotic
  maps via state-mapping networks, IEEE Transactions on Circuits and Systems I:
  Regular Papers 66~(6) (2019) 2322--2335.
\newblock \href {http://dx.doi.org/10.1109/TCSI.2018.2888688}
  {\path{doi:10.1109/TCSI.2018.2888688}}.

\bibitem{Li:Cryptanalysis:IM2018}
M.~Li, H.~Fan, Y.~Xiang, Y.~Li, Y.~Zhang, Cryptanalysis and improvement of a
  chaotic image encryption by first-order time-delay system, IEEE Multimedia
  25~(3) (2018) 92--101.
\newblock \href {http://dx.doi.org/10.1109/MMUL.2018.112142439}
  {\path{doi:10.1109/MMUL.2018.112142439}}.

\bibitem{F:nonlinear:ND18}
F.~\"Ozkaynak, Brief review on application of nonlinear dynamics in image
  encryption, Nonlinear Dynamics 92 (2018) 305--313.
\newblock \href {http://dx.doi.org/10.1007/s11071-018-4056-x}
  {\path{doi:10.1007/s11071-018-4056-x}}.

\bibitem{cxz:RT-enhanced:IEEEA18}
C.~Zhu, K.~Sun, Cryptanalyzing and improving a novel color image encryption
  algorithm using {RT}-enhanced chaotic tent maps, IEEE Access 6 (2018)
  18759--18770.
\newblock \href {http://dx.doi.org/10.1109/ACCESS.2018.2817600}
  {\path{doi:10.1109/ACCESS.2018.2817600}}.

\bibitem{Latif:Quantum:TNSM20}
A.~A.~A. El-Latif, B.~Abd-El-Atty, W.~Mazurczyk, C.~Fung, S.~E.
  Venegas-Andraca, Secure data encryption based on quantum walks for {5G}
  internet of things scenario, IEEE Transactions on Network and Service
  Management 17~(1) (2020) 191--217.
\newblock \href {http://dx.doi.org/10.1109/TNSM.2020.2969863}
  {\path{doi:10.1109/TNSM.2020.2969863}}.

\bibitem{lfl:chaotic-maps:IET17}
L.~Liu, S.~Hao, J.~Lin, Z.~Wang, Image block encryption algorithm based on
  chaotic maps, IET Signal Processing 12 (2017) 22--30.
\newblock \href {http://dx.doi.org/10.1049/iet-spr.2016.0584}
  {\path{doi:10.1049/iet-spr.2016.0584}}.

\bibitem{cql:Inf-Entropy:IEEE18}
C.~Li, B.~Feng, J.~L\"u, Cryptanalysis of a chaotic image encryption algorithm
  based on information entropy, IEEE Access 6 (2018) 75834--75842.
\newblock \href {http://dx.doi.org/10.1109/ACCESS.2018.2883690}
  {\path{doi:10.1109/ACCESS.2018.2883690}}.

\bibitem{Lcq:Optimal:SP11}
C.~Li, K.-T. Lo, Optimal quantitative cryptanalysis of permutation-only
  multimedia ciphers against plaintext attacks, Signal Processing 91~(4) (2011)
  949--954.
\newblock \href {http://dx.doi.org/10.1016/j.sigpro.2010.09.014}
  {\path{doi:10.1016/j.sigpro.2010.09.014}}.

\bibitem{fhj:Cryp-APFM:SP18}
H.~Fan, M.~Li, D.~Liu, E.~Zhang, Cryptanalysis of a color image encryption
  using chaotic {APFM} nonlinear adaptive filter, Signal Processing 143 (2018)
  28--41.
\newblock \href {http://dx.doi.org/10.1016/j.sigpro.2017.08.018}
  {\path{doi:10.1016/j.sigpro.2017.08.018}}.

\bibitem{cql:hierarchical:SP16}
C.~Li, Cracking a hierarchical chaotic image encryption algorithm based on
  permutation, Signal Processing 118 (2016) 203--210.
\newblock \href {http://dx.doi.org/10.1016/j.sigpro.2015.07.008}
  {\path{doi:10.1016/j.sigpro.2015.07.008}}.

\bibitem{Preishuber:motivation:TIFS2018}
M.~Preishuber, T.~Huetter, S.~Katzenbeisser, A.~Uhl, Depreciating motivation
  and empirical security analysis of chaos-based image and video encryption,
  IEEE Transactions on Information Forensics and Security 13~(9) (2018)
  2137--2150.
\newblock \href {http://dx.doi.org/10.1109/TIFS.2018.2812080}
  {\path{doi:10.1109/TIFS.2018.2812080}}.

\bibitem{cqli:cat:2019}
C.~Li, K.~Tan, B.~Feng, J.~L\"u, The graph structure of the generalized
  discrete {A}rnold's {C}at map, https://arxiv.org/abs/1712.07905.

\bibitem{Fan:period:ND19}
C.~Fan, Q.~Ding, Analysing the dynamics of digital chaotic maps via a new
  period search algorithm, Nonlinear Dynamics 97~(1) (2019) 831--841.
\newblock \href {http://dx.doi.org/10.1007/s11071-019-05015-4}
  {\path{doi:10.1007/s11071-019-05015-4}}.

\end{thebibliography}

\end{document}